\documentclass[superscriptaddress,twocolumn,prb,showpacs]{revtex4}
\usepackage{graphics}

\begin{document}

\title{Scaling determination of the nonlinear $I$-$V$ characteristics
for 2D superconducting networks}

\author{Petter \surname{Minnhagen}}
\affiliation{NORDITA, Blegdamsvej 17, Dk 2100, Copenhagen,
Denmark}
\affiliation{Department of Theoretical Physics, Ume{\aa}
University, 901 87 Ume{\aa}, Sweden}

\author{Beom Jun \surname{Kim}}
\affiliation{Department of Molecular Science and Technology, Ajou
  University, Suwon 442-749, Korea}
\author{Andreas  \surname{Gr\"{o}nlund}}
\affiliation{Department of Theoretical Physics, Ume{\aa}
University, 901 87 Ume{\aa}, Sweden}


\begin{abstract}
It is shown from computer simulations that the current-voltage
($I$-$V$) characteristics for the two-dimensional $XY$ model with
resistively-shunted Josephson junction  dynamics and Monte Carlo
dynamics obeys a finite-size scaling form from which the nonlinear
$I$-$V$ exponent $a$ can be determined to good precision. This
determination supports the conclusion $a=z+1$, where $z$ is the
dynamic critical exponent. The results are discussed in the light
of the contrary conclusion reached by Tang and Chen [Phys. Rev. B
{\bf 67}, 024508 (2003)] and the possibility of a breakdown of
scaling suggested by Bormann [Phys. Rev. Lett. {\bf 78}, 4324
(1997)].
\end{abstract}

\maketitle

\section{Introduction}
The physics of the static Kosterlitz-Thouless (KT) transition~\cite{ktb}
for quasi-two-dimensional (2D) superconductors is well understood
in terms of vortex pair unbinding, since long
time.~\cite{minnhagen_rev} However, in spite of this and in spite
of the large interest in quasi-2D superconductors over recent
years spurred by the event of high-$T_c$
superconductors,~\cite{blatter_rev,minnhagen_ferm} the dynamics of
2D vortices in the vicinity of the KT transition is still not
completely settled. In the present paper we address the specific
question of the value of the nonlinear $I$-$V$ exponent $a$ in the
low temperature superconducting phase, which has been brought up
by Tang {\it  et al.} in two recent papers.~\cite{tang1,tang2}

The first attempt to describe the dynamics close to the
KT transition was made by Ambegaokar-Halperin-Nelson-Siggia
(AHNS).~\cite{ahns}
This attempt was based on the phenomenological assumption that the
vortices could be separated into two distinct categories, i.e.,
free individual vortices and bound pairs of vortices and
antivortices. This reasoning led to the AHNS value $a=a_{\rm
AHNS}$ for the nonlinear $I$-$V$ exponent in $V\propto I^{a}$ in the
low temperature phase.~\cite{ahns} An alternative value was later suggested by
Minnhagen {\it et al.}~\cite{minnhagen_scale} based on the observation
that the low temperature phase is quasicritical and as a
consequence the exponent $a$ should follow from critical scaling.
This gave the exponent $a=a_{\rm scale}=z+1$ where $z$ is the
dynamic critical exponent. However, these two alternative values
are different $a_{\rm scale}>a_{\rm AHNS}$ for $T<T_{\rm KT}$. The
question is then which one is correct. This is the question
discussed in the present paper.

A series of attempts to settle this issue was based on numerical
simulations for various models displaying a KT transition in 2D: The
Coulomb gas with Langevin dynamics,~\cite{holmlund} the
lattice Coulomb gas with Monte Carlo (MC) dynamics,~\cite{weber} the
$XY$ model with resistively-shunted Josephson junction (RSJ)
dynamics.~\cite{minnhagen_scale,kmo} These simulations all
suggested that $a_{\rm scale}$ was the correct exponent.

However, for the 2D $XY$ model with RSJ dynamics in
Ref.~\onlinecite{simkin}, Simkin and Kosterlitz argued that their
data were more consistent with $a_{\rm AHNS}$. This discrepancy
between the result of this simulation and the result obtained in the
others cited above reflects the general difficulty in the
determination of $a$: In a log-log plot the power law form of the
nonlinear $I$-$V$ characteristics should give a straight line with
the slope $a$. However, this form is only valid for small enough
$I$ and the smaller the $I$ the larger system size is needed in
order to get a size independent results. Thus great care is required
to ensure that the data used in the determination of the slope are
really size independent, which in turn makes the simulations
demanding.

The experimental determination of $a$ is faced with the same type
of finite size problems, as recently discussed in e.g.
Ref.\onlinecite{strachan}. However, in order to experimentally
distinguish between the two different predictions of $a$ one needs
in addition to analyze the temperature dependence of $a$ as
described in Ref.\onlinecite{minnhagen_ferm}. This makes the
experimental route to settle the issue somewhat difficult.

On the other hand, the size dependence of the data can also be
turned into an advantage when determining $a$ from simulations:\cite{kmo}
The logic here is that by carefully choosing the boundary condition the data
should obey a size scaling from which $a$ can be determined {\em
provided} the scaling assumption leading to $a_{\rm scale}$ is
indeed correct. This strategy was used in Ref.~\onlinecite{kmo} and
provided strong evidence in favor of the critical scaling and
$a_{\rm scale}$, as will be discussed in more detail in the
present paper.

In an attempt to resolve the issue of the two different results
for the value of $a$, Bormann in Ref.~\onlinecite{bormann}
reinvestigated the AHNS reasoning and concluded that it contains
both of the results: For small enough current $I$ the result
$a=a_{\rm AHNS}$ should be correct but as $I$ is increased there
should be a crossover to a distinct region where $a=a_{\rm
scale}$. In the $a=a_{\rm scale}$ regime the scaling should hold,
so according to this analysis one should see a breakdown of
scaling into an AHNS regime for small enough $I$. As found in
Ref.~\onlinecite{kmo} and which will be further discussed here, the
scaling assumption holds for all the data obtained and there is no
sign of a crossover to an AHNS regime.

The most recent attempt to settle this issue is by Chen, Tang, and
Tong in Refs.~\onlinecite{tang1,tang2}. They again try to estimate
the nonlinear $I$-$V$ exponent $a$, ($V\propto I^a$) below the
KT transition for a 2D Josephson junction array from
the slope of $\ln V $ versus $\ln I$ for small $I$. The basic
claim made is that an anomalous finite-size effect for small $I$
(meaning that in a certain parameter range the voltage for a fixed
small $I$ increases with size instead of decreases) gives an
overestimation of $a$. As a consequence it was concluded that $a$
is in better agreement with the AHNS prediction~\cite{ahns}instead
of the dynamic scaling prediction~\cite{minnhagen_scale} concluded
in Ref.~\onlinecite{kmo,choi}. However, as shown in
Ref.~\onlinecite{kmo}, such an anomalous finite-size effect is a
feature consistent with and emerging from the dynamical scaling
and does consequently not affect the reliability of a
determination based on finite-size scaling.

Our strategy to settle the issue is based on the observation that
the dynamical scaling alternative is very amenable to testing by
computer simulations.~\cite{kmo} This is because the dynamical
scaling makes direct predictions of the data obtained for finite
size systems and does thus not hinge on any estimate of the
asymptotic slope in the limit of small $I$ and large size $L$.
Using this approach, a result consistent with AHNS requires that
the dynamical scaling does in fact {\em fail} to describe the
data. Alternatively, if one wants to verify a crossover to AHNS
from a scaling regime, as suggested by Bormann in
Ref.~\onlinecite{bormann}, then one needs to demonstrate that the
dynamical scaling breaks down as one passes over into the AHNS
regime. As shown in the present paper and in Ref.~\onlinecite{kmo},
direct tests of the dynamical scaling through simulations of the
2D $XY$ model with RSJ dynamics and MC dynamics give excellent
agreement with dynamical scaling without any sign of the breakdown
required for the AHNS alternative to be valid.

The content of the paper is as follows: In Sec. II we briefly
recapitulate the finite-size dynamical scaling and in Sec. III
the results of the simulations and the data analysis are presented.
In Sec. IV we discuss our results in the context of other
attempts to settle the issue and Sec. V, finally, contains some
concluding remarks.

\section{The dynamical scaling approach}
The method used to determine $a$ directly from finite-size scaling
is described in Ref.~\onlinecite{kmo}. It is based on the usual
scaling form by Fischer {\it et al}. in Ref.~\onlinecite{fischer} adopted to
two dimensions and to finite-size scaling at criticality. Since the
low temperature phase is quasicritical the scaling form applies
at and below the KT transition.~\cite{kmo,dorsey} In order to take
maximum advantage of the size scaling we reduce the number of
large scales to one by considering a quadratic system with side
$L$. The size scaling form is then given by~\cite{kmo}
\begin{equation}\label{scale}
v=i^{z+1}\left[ \frac{f(Li)}{Li}\right]^z
\end{equation}
where $v=V/L$ is the voltage per length across the sample, $i=I/L$
is the current density, $z$ is dynamic critical exponent and
$f(x)$ is a scaling function which goes to a positive constant for
small $x$ and is proportional to $x$ for large $x$. This means
that $v\propto i^{z+1}$ for a given small enough $i$ in the limit
of large $L$. Consequently, the scaling prediction for the
nonlinear $I$-$V$ exponent $a$ ($V\propto I^a$) is $a=z+1$.

The use of this scaling approach has several advantages when
trying to determine the exponent $a$ in model simulations: First
of all, it does not hinge on the accuracy to calculate the small
$i$-limit of voltages where at the same time $Li>>1$, which is
notoriously difficult. Secondly, the existence of a scaling
function $f(x)$ can be determined from a data collapse using data
from all system sizes simultaneously. Thirdly, $z$ can be related
to and obtained from the equilibrium properties of the
system~\cite{kmo}. This gives an independent consistency check on
the scaling given by Eq.(\ref{scale}), since this $z$-value has to
agree with the one obtained directly from the data collapse.

\section{Results for the 2D $XY$ model}
Our main results are for the 2D $XY$ model with RSJ dynamics and
finite temperatures $T$. This model undergoes a KT transition at
$T_c\approx 0.89$ ($T$ is measured in units of the Josephson
coupling and current in terms of the critical current $i_c$ of a
single Josephson junction). The method used is described in
Ref.~\onlinecite{kmo}: We use a fluctuating twist boundary condition (FTBC)
and a square lattice with size $L$. The data presented here are
well converged data for sizes up to $L=256$. We here analyze data
obtained in the low temperature quasicritical phase somewhat
below $T_c$.

\begin{figure}
\centering{\resizebox*{!}{5.8cm}{\includegraphics{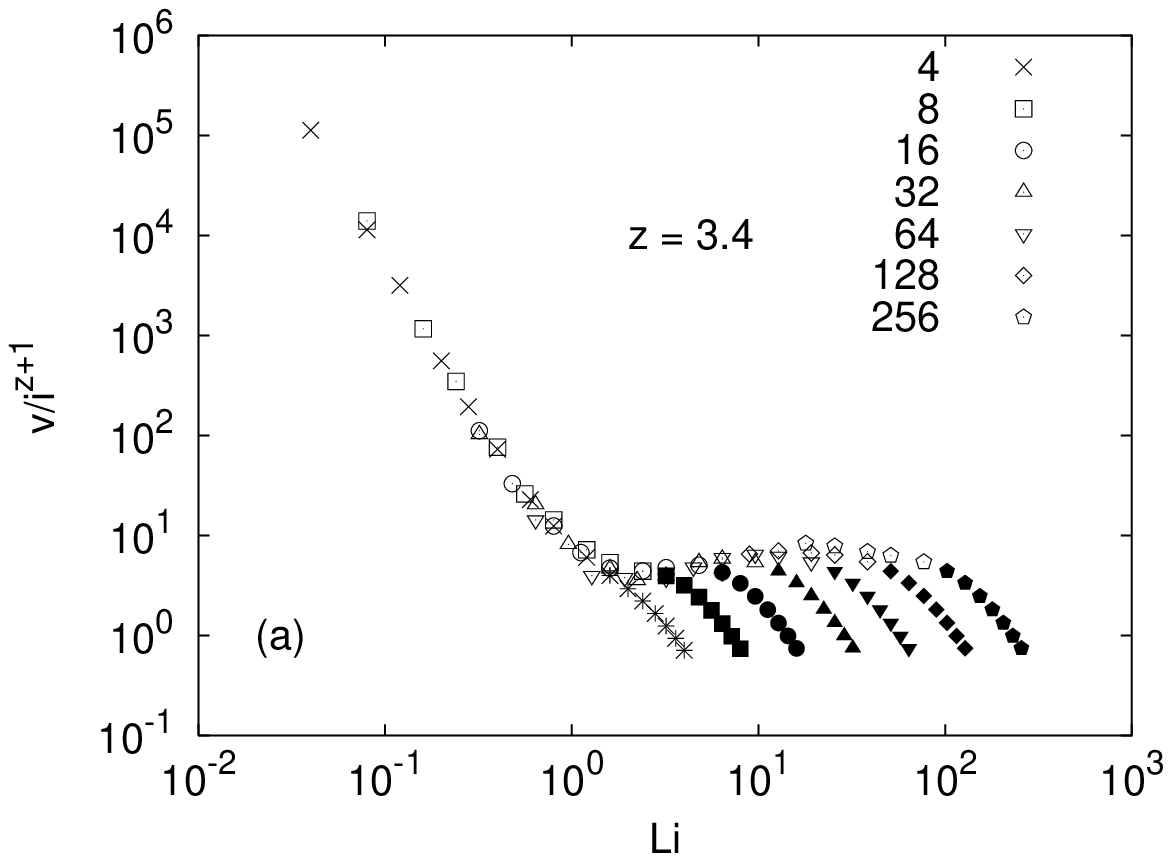}}}
\centering{\resizebox*{!}{5.8cm}{\includegraphics{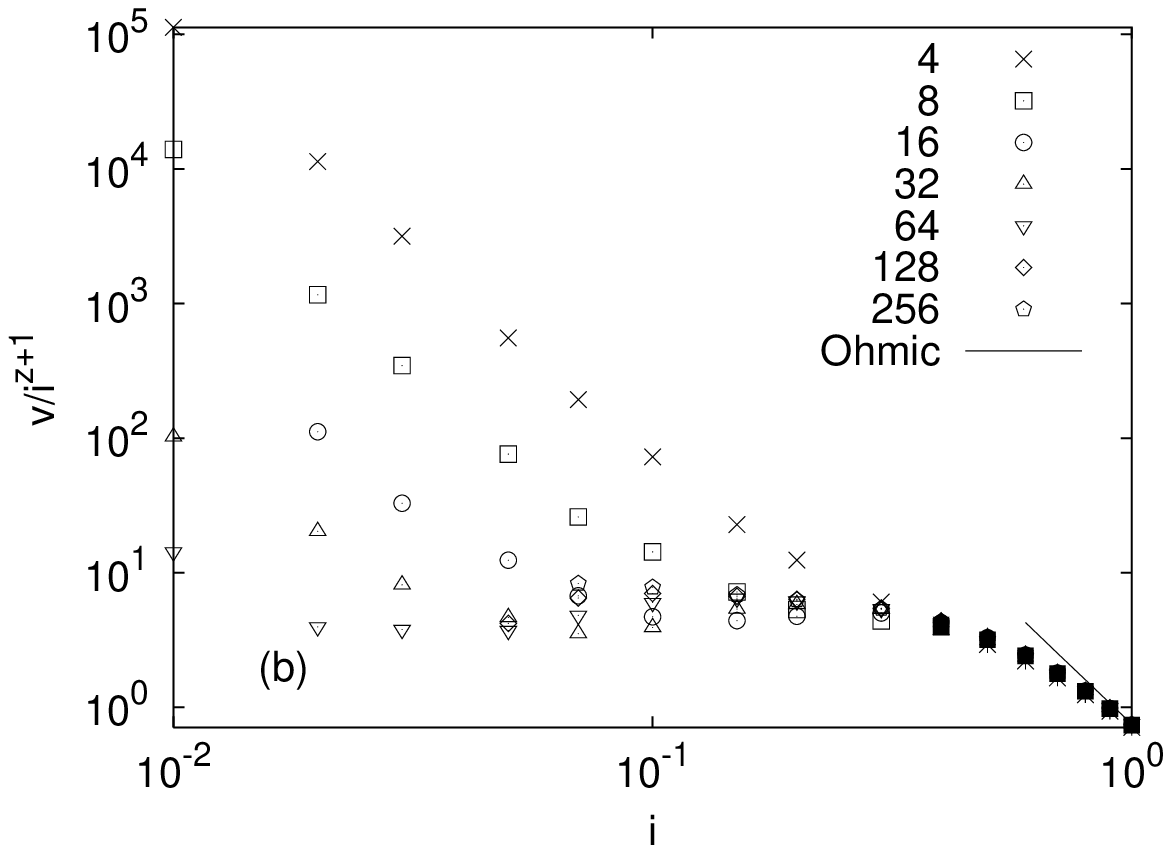}}}
\caption{a) Demonstration of the size scaling for the 2D XY model
with RSJ dynamics for sizes $L=4$ to $L=256$. The size scaling
function $F(x)=[f(x)/x]^z$ (compare Eq.\ref{scale}) is obtained by
plotting $v/i^{z+1}$ against $Li$. The value of $z=3.4$ used in
this plot is independently obtained from equilibrium quantities
(see Ref.\onlinecite{kmo}). The data fall on the scaling curve
(open symbols) except for the data corresponding to the highest
currents ($i\geq 0.4$, filled symbols). b) The same data is
plotted as $v/i^{z+1}$  against $i$. Now all the high current data
($i\geq 0.4$, filled symbols) collapse onto a single curve. This
means that the high current data to good approximation are size
independent and belongs to the crossover region towards the ohmic
limit which is controlled by the critical current for a single
Josephson junction $i_c=1$. The ohmic limit corresponds to the
straight line.} \label{Fig1}
\end{figure}

In Fig.~1 we present results for $T=0.8$ in the alternative
scaling form $v/i^{z+1}=F(Li)$ where $F(x)$ is related to $f(x)$
in Eq.~(\ref{scale}) by $F(x)=\left[ f(x)/x\right]^z$. In this
plot we use the value $z=3.4$ obtained from equilibrium
simulations using the relation~\cite{minnhagen_scale}
$z=1/(\tilde{\epsilon}T^{CG})-2$ with the dielectric constant
$\tilde\epsilon$ and the Coulomb gas temperature $T^{CG}$, as
described in Ref.~\onlinecite{kmo} and given in Table I of
Ref.~\onlinecite{kmo}.~\cite{value} As seen from Fig.~1(a) a very
good data collapse is obtained. At higher values of $Li$ there is
a notable systematic deviation from scaling. In order to determine
the cause of this deviation the same data are plotted in Fig.~1(b)
as a function of $i$. As seen the data which in Fig.~1(a) deviate
from the scaling curve [filled symbols in Fig.~1(a) and (b)] now
instead collapse and furthermore the collapse for all sizes starts
approximately at the same value of $i\approx 0.3$. This means that
the data for values $i>0.3$ are independent of $L$. The Josephson
junctions become completely resistive for $i=i_c$ with a
resistance given by normal state $R_N$ of the junction and which
also means that $V=R_NI$ for the complete array. This large
current relation is given by the full line in Fig.~1(b). It means
that the deviation from the scaling curve in Fig.~1(a) is just the
trivial crossover to the normal state resistance which always
occurs for large enough $i$ because the small scale $1/i_c$
becomes relevant and breaks the scaling when $i$ becomes of the
same order as $i_c$. The data in Fig.~1(a) show no deviation from
finite-size scaling apart from the trivial crossover to normal
state resistance as $i$ approaches $i_c$.

\begin{figure}
\centering{\resizebox*{!}{5.8cm}{\rotatebox{270}{\includegraphics{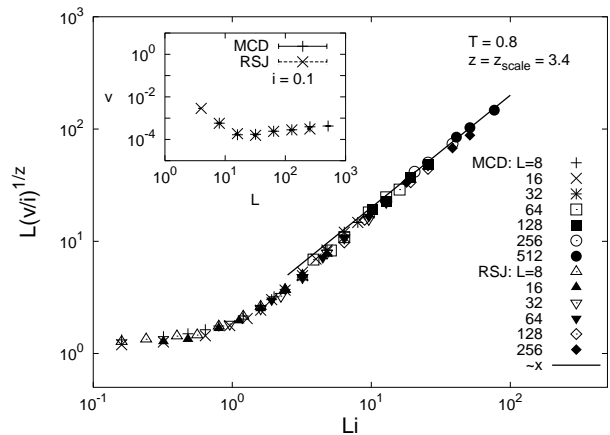}}}}
\caption{Joint demonstration of size scaling for the 2D XY model
with RSJ and MC dynamics. The size scaling function $f(x)$ in
Eq.(\ref{scale}) is obtained by plotting $L(v/i)^{1/z}$ against
$Li$. The highest current data belonging to the cross region are
not included. The sizes spanned are for RSJ $L=8-256$ and for MC
$L=8-512$. The independently obtained value $z=3.4$ is used (see
Fig.1). The straight line corresponds to the large $x$ limit $f(x)
\propto x$ whereas the small $x$ limit is given by $f(x)=const$.
The inset shows the size dependence for a fixed current $i=0.1$.
One notes that $v$ in this case increases for large $L$. These
type of behavior was in Ref.\onlinecite{tang1} termed anomalous
size dependence. However, it does only reflect the shape of the
size scaling function and not any distinct anomalous physics.}
\label{Fig2}
\end{figure}

Figure~2 shows the scaling function $f(Li)$ in Eq.~(\ref{scale})
obtained from the same data. In addition, we have in this figure
also included data from the 2D $XY$ model with Monte Carlo (MC)
dynamics. As demonstrated in Ref.~\onlinecite{katya}, MC dynamics
gives for the same $T$ the same $I$-$V$ characteristics up to a
constant factor and the same method as in Ref.~\onlinecite{katya}
is used here. The MC data are also for $T=0.8$ but also include
the larger size $L=512$. The data for $i$ in the crossover towards
$v\propto i$ have been excluded in Fig.~2. As seen both sets of
data collapse on the scaling curve with $z=3.4$. The full line
($\propto Li$) corresponds to $v\propto i^{z+1}$. Note that the
data for a given size $L$ which fall on this line have the scaling
slope $a_{\rm scale}$ in a log-log plot of $v$ versus $i$. This
fact also gives an idea of the difficulty with a direct
determination for a single system size: The larger system the
fewer points can in practice be obtained. So although the three
data points for $L=512$ fall nicely on the scaling function a
determination of $a$ based on these three points alone would not
have the same reliability.

The inset in Fig.~2 shows how the voltage $v$ depends on the size
$L$ for a given fixed current. As seen the voltage for this fixed
current increases for larger $L$. This is just a reflection of the
fact that the scaling curve falls below the full line in Fig.~2
for $Li$ between approximately 1 and 10. This means that the
increase of the voltage with increasing $L$ is a property of the
scaling regime where the behavior is controlled by a single
relevant large scale. Consequently, it is not connected to the
appearance of a second relevant scale, as suggested in Ref.6 in
case of the seemingly very similar increase of the voltage with
increasing $L$ found in this paper.

\section{Comparisons and discussion}
Our main result from the 2D $XY$ model with RSJ and MC dynamics is
that the data collapse on a scaling curve for a certain value of
$z$. This value of $z$ is in agreement with the scaling prediction
which connects the value of $z$ to equilibrium
quantities.~\cite{minnhagen_scale,kmo} The data regime at higher
current which do not fall on the scaling curve is controlled by
the Josephson junction critical current $i_c$ and belong to the
crossover to the normal state linear relation $V=R_NI$. The data
show {\em no} crossover to an AHNS behavior.

According to Bormann in Ref.~\onlinecite{bormann} such a crossover
should be expected for small enough $i$. Thus, according to this
theory, AHNS should be valid for low enough $i$ in the limit of
large $L$ and there should be a crossover to the scaling result
for larger $i$. As $i$ approaches the critical Josephson junction
current there should be a second crossover towards the normal
state linear $I$-$V$ characteristics. As seen from Fig.~2 the data
give evidence of a crossover from the scaling regime to the linear
Ohmic regime but there is no sign of a crossover towards the AHNS
regime for smaller $i$. In Fig.~2 such a crossover would mean that
the data for larger $L$ and smaller $i$ would fall above the
scaling curve. The crossover current $i_{\rm cross}$ is
proportional to $1/n^{1/(z/2-1)}$ where $n$ is the vortex density.
An attempt towards a more quantitative estimate of the crossover
is given in Fig.~1 of Ref.~\onlinecite{bormann}: Our data for
$i=0.01$ correspond to a point in this figure $(x,y)=(z+2,
i^{-1})=(6.4, 10^2)$ which should be far inside the AHNS regime.
In fact all the data up to the crossover to the linear regime (at
$i\approx 0.3$) should, according to this estimate, belong to the
AHNS regime. However, our data show no deviation from scaling for
any size or current, indicating that all the data are instead in
the scaling regime and that there is no crossover to an AHNS
regime at small currents. However, one can, of course, not
entirely rule out that such a crossover might exist for even lower
currents and larger sizes than could be reached in the
simulations.

Tang {\it et al.} in Ref.~\onlinecite{tang1,tang2} use a
rectangular sample with sides $L_x$ and $L_y$ and the current in
the $x$ direction. The idea is to use a small ratio $L_y/L_x$ in
order to minimize the dependence on the scale $L_x$ and they in
practice use $0.004\leq L_y/L_x\leq 0.25$ corresponding to $L_y=8$
and $L_y=512$, respectively. The current injection method used
introduces a nonuniform vortex density in the current direction.
However this is compensated by skipping a boundary region with the
length $b$ at both boundaries when measuring the voltage. In
contrast the use of FTBC is designed to make the vortex density
uniform. Nevertheless, it was shown in Ref.~\onlinecite{choi} that
the method adopted by Tang {\it et al.}~\cite{tang1,tang2} for PBC
in the transverse direction gives the same result as the FTBC
method provided care is taken to avoid any influence from the two
additional length scales $L_x$ and $b$. Ref.~\onlinecite{choi}
confirms the scaling prediction for sizes up to $L_y=64$. Tang
{\it et al.}~\cite{tang2} argues that in the case of a periodic
boundary condition (PBC) in the direction transverse to the
current one should get a crossover from a scaling regime at {\em
lower} currents towards an AHNS regime at {\em higher}. Note that
this is precisely the {\em opposite} to the Bormann
prediction.~\cite{bormann} According to Tang {\it  et al.} such a
crossover is supported by their Fig.~6 in Ref.~\onlinecite{tang2}.
This figure for $T=0.8$ corresponds precisely to our Fig.~1(a) for
the same transverse sizes. The scaling is clearly visible although
the quality in their Fig.~6 is not as good as in our
Fig.~1a.\cite{magnetude} However, whereas our method is
specifically designed to test the scaling and only introduces a
single scale $L$, the method by Tang {\it et al.}~\cite{tang2}
have three large scales and in particular the fact that the ratio
$L_y/L_x$ varies from 0.004 to 0.25 instead of being constant,
might well cause deviations from scaling. In addition, there is a
crossover for higher currents towards the linear $I$-$V$
characteristics. However, whereas, as shown above, the crossover
in our case can unambiguously be attributed to the trivial
crossover to the linear regime starting at around $i\approx 0.3$
for all sizes [compare Fig.~1(b)] the situation in Fig.~6 of Tang
{\it et al.}~\cite{tang2} is slightly different. At small values
of $L_y$ the crossover comes at roughly the same value $i\approx
0.3$ (up to $L\approx 32$) and then it decreases to roughly 0.06
at $L_y=512$. This decrease of the crossover current with $L_y$
indicates that there is another length scale in the problem. Tang
{\it et al.} suggest that this new length scale is associated with
the vortex physics and is given $L_r\propto i^{-(z+2)/4}$ and that
the crossover is to an AHNS regime occurring for $L_y>L_r$. The
problem with this interpretation is, in the light of Fig.~1(a) and
(b), that there is {\em no} such size dependent crossover for a
square lattice. Since the bulk properties of vortex physics cannot
depend on the shape of the sample, the complete absence of the
appearance of an additional vortex length in Fig.~1(a) and (b)
strongly suggests that the explanation offered by Tang {\it et
al.}~\cite{tang2} cannot be the correct one. An obvious candidate
is instead the additional ratio $L_y/L_x$ introduced by Tang {\it
et al.} and which is not kept fixed when $L_y$ is varied, as
required by a proper scaling analysis. The point to note is that
it is not the absolute value of $L_x$ per se that matters, but the
fact that the ratio $L_y/L_x$ has to be constant, and as pointed
out above it varies from very small to 0.25 in
Ref.\onlinecite{tang2}. This could explain both why the quality of
the size scaling is not as good as for the scaling with fixed
$L_y/L_x=1$ shown in our Fig.~1a\cite{magnetude}, as well as the
size dependence of the crossover current.

\begin{figure}
\centering{\resizebox*{!}{5.8cm}{\includegraphics{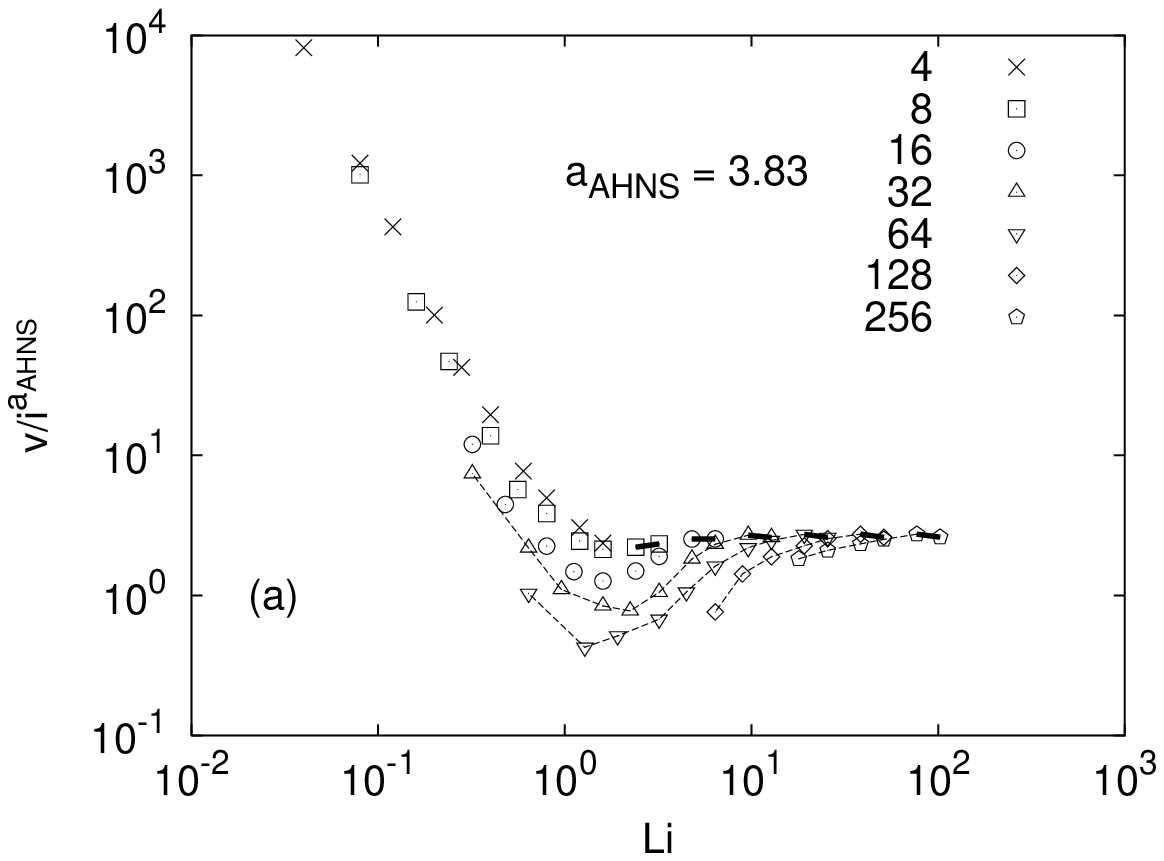}}}
\centering{\resizebox*{!}{5.8cm}{\includegraphics{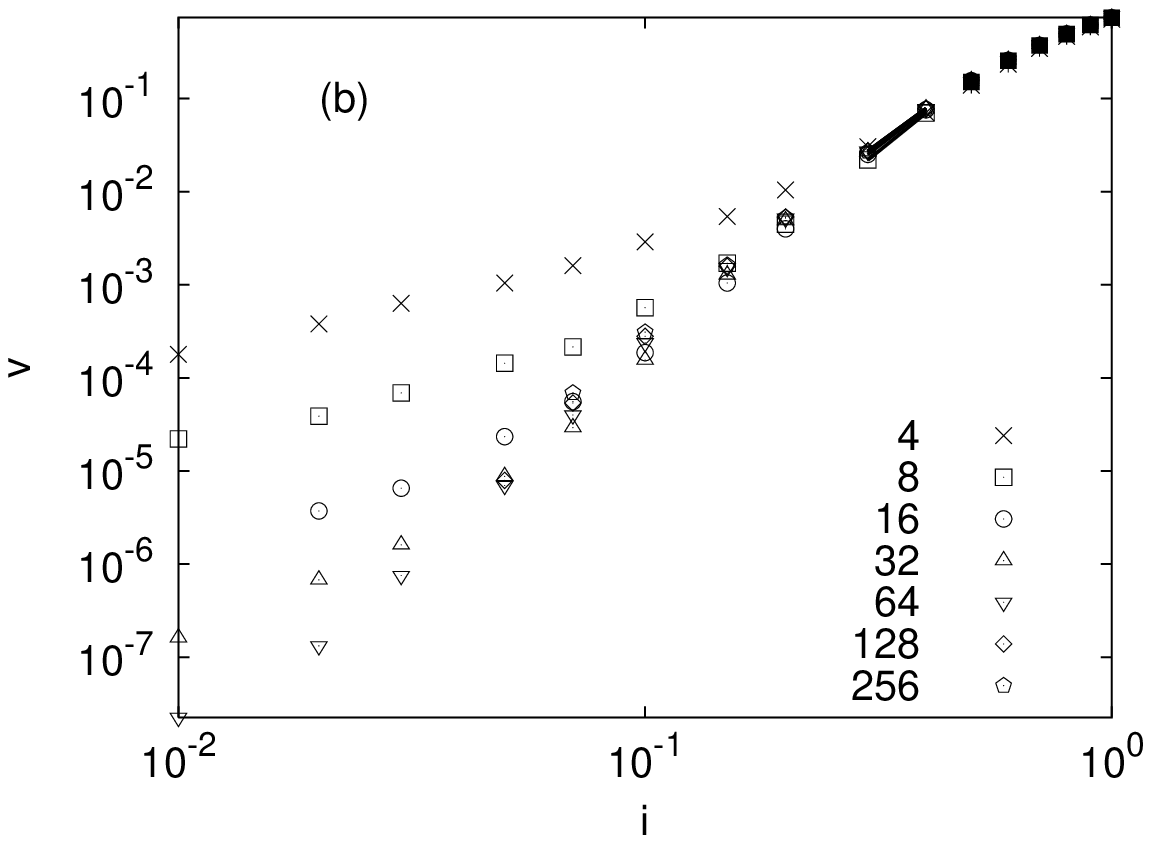}}}
\caption{ a) The data for the 2D XY model with RSJ dynamics are
plotted as $v/i^{a_{{\rm AHNS}}}$ against $Li$ with $a_{{\rm
AHNS}}=3.83$(see Ref.\onlinecite{tang2}, just as in Fig.8 of this
reference the data points for $i>0.4$ are excluded in the plot).
Part of the data seemingly fall on a horizontal line (data pairs
connected with lines) indicating that $v\propto i^{a_{{\rm
AHNS}}}$. This is a false conclusion because the pairs only
constitute a repetition of the same two size converged data
points. These two data points, which correspond to $i=0.3$ and
0.4, are shown in Fig.3b. The point to note in a) is that the data
for the larger sizes and smaller currents deviate towards lower
values (dashed curves are guides to the eye). Thus this plot with
$v/i^{a_{{\rm AHNS}}}$ does not correspond to any true size
scaling function, in contrast to Fig.1a. b) Plot of $v$ versus $i$
for the same data as in a). There are always two data points in
the cross over region towards the ohmic limit which can be joined
together with a line of slope $a_{{\rm AHNS}}$ (in the present
case the two points with $i=0.3$ and 0.4). The construction in
Fig.3a repeats these two data points into a horizontal line.
Fig.3a and b should be compared to Fig.8 and 7 in
Ref.\onlinecite{tang2}, respectively.} \label{Fig3}
\end{figure}

Tang {\it et al.}~\cite{tang2} also obtained the data for an open
boundary conditions (OBC) in the transverse direction. The
difference with the PBC result is the different finite-size
dependence of the voltage. The open boundary causes stronger
boundary effects and thus changes the details of the size
convergence. The data are plotted as $v/i^{a_{\rm AHNS}}$ versus
$iL_y$ in Fig.~8 in Ref.~\onlinecite{tang2}. In Fig.~3(a) our data
for FTBC and square lattice are plotted in the same way. As seen
in Fig.~3(a) the data over a large region apparently fall on a
horizontal line. This line corresponds to $v\propto i^{a_{\rm
AHNS}}$ and consequently one might be tempted to interpret this as
evidence of an AHNS exponent. The fallacy here is that the data
which constitute this line are the size converged data in the
crossover region towards $i_c$ as shown in Fig.~3(b): The two data
points connected with a line in Fig.~3(b) forms a spurious AHNS
line by repetition in Fig.~3(a). When going from the steeper slope
in the scaling region through the crossover to the linear $I$-$V$
dependence there will always be some current region where the
slope has the AHNS value. However, this continuous crossover to a
linear $I$-$V$ characteristics has nothing to do with the AHNS
vortex physics. The important thing to notice in Fig.~3(a) is that
the scaling systematically fails for the larger sizes at lower $i$
where the data for each size deviate to lower values with
decreasing $i$ for a fixed size. The same deviation is observed in
Fig.~8 in Ref.~\onlinecite{tang2} where the data on the horizontal
line are a repetition of the two data points in Fig.~7 from
Ref.~\onlinecite{tang2} for the current sizes $i=0.3$ and 0.4
(same currents as in Fig.3a), respectively. Thus for the larger
system sizes, which are relevant for the bulk properties of vortex
physics, the change of boundary condition to OBC makes little
difference. This is of course expected because the boundary
condition should not matter at all for large enough samples. For
smaller samples the results are different, as apparent when
comparing our Fig.~3(a) with Fig.~8 in Ref.~\onlinecite{tang2}.
This is also expected because here the fluctuation associated with
the open boundary increases the voltage for a given $i$ relative
to the PBC boundary. However, the vortex physics in the bulk
remains the same and it is these properties which dominate for
large samples.

To summarize: Neither the shape of the sample nor the details of
the boundary condition can change the vortex physics in the bulk
for large enough systems. We find that there is {\em no}
compelling evidence that the data by Tang {\it et
al.}~\cite{tang2} are in contradiction with this statement. Proper
scaling analysis with a fixed ratio $L_y/L_x$, as shown in the
present paper, gives {\em no} evidence of an extra scale in
addition to $L$ and $i_c$.

\section{Concluding remarks}

As discussed in the present paper the $I$-$V$ characteristics for
the 2D $XY$ model with RSJ and MC dynamics obey size scaling for a
quadratic sample (scaling in $L_y$ with fixed at  $L_y/L_x=1$).
The fact that the data obey size scaling means that $a=z+1$ where
$z$ is the dynamic critical exponent. The exponent $z$ can be
calculated from the equilibrium properties of the system which
together with the relation $a=z+1$ gives the scaling prediction
for $a$. As shown this scaling prediction is in agreement with the
$z$ value determined from the data collapse. The only deviation
from the scaling can, as shown here, be linked to the trivial high
current crossover to the linear $I$-$V$ relation.

A prediction for $a$ which is different from the scaling
prediction can only be correct if the scaling breaks down. As
shown in the present paper, there is no such breakdown in the
scaling except for the one at high currents associated with the
critical current of a single Josephson junction. This does in
principle not rule out the possibility that a breakdown could
occur at sizes larger and currents smaller than reached in the
simulations. However, whereas the scaling prediction has been
verified over a large parameter range, the breakdown of scaling,
which would have to occur if the AHNS prediction was correct in
the ultimate small current limit, has not been supported by any
simulation data so far.

In our opinion, the only convincing way to verify that AHNS, or
some other alternative prediction is correct in some parameter
regime, is to demonstrate that the scaling breaks down in the same
regime.

\section*{Acknowledgements}
Support from the Swedish Research Council is gratefully
acknowledged. B.J.K acknowledges the support from the Korea
Science and Engineering Foundation through Grant No.\
R14-2002-062-01000-0, and the support from Ajou University.


\begin{thebibliography}{99}
\bibitem{value} In order to obtain optimal consistency the
equilibrium value of $z$ is calculated with the same
boundary condition as the dynamical calculation. Note that for
this boundary condition the helicity modulus is identically zero.
See Ref.~\onlinecite{kmo}.
\end{thebibliography}
\end{document}